# Investigation of the characteristics of the dual-band electromagnetic induction transparent-like terahertz spectrum in a grating-like structure

Chaoying Zhao[1,2,] Jiahao Hu[1]
[1] College of Science, Hangzhou Dianzi University, Zhejiang 310018, China
[2] State Key Laboratory of Quantum Optics and Quantum Optics Devices, Institute of Opto-Electronics, Shanxi University, Taiyuan 030006, China
**E-mail**:
 zchy49@163.com

**Abstract**

In this paper, a new structure is proposed which is expected to realize dual-band electromagnetically induced transparency (EIT)-like effect in terahertz spectrum. It can be regarded as a metamaterial of grating-like elements composed of dual-band U-shaped resonators. The near-field coupling between bright modes can produce EIT-like effect. After optimizing the parameters, our numerical simulated results are in good agreement with the theoretical analysis. The EIT-like effect can significantly reduce the group speed near the transparent window, may gain more significant potential applications in slow-light transmission and optical storage.

**1. Introduction**

In 1990, Harri first discovered the electromagnetic induction transparency(EIT) phenomenon in atomic vapor [1]. EIT is a coherent process, in which the opaque atomic medium can produce a transparent window in a narrow region in the wide absorption band through the quantum interference of the pump light and the probe light at different energy level transitions[2,3]. The EIT effect occurs when the atomic medium accompany a strong dispersion, which can significantly reduce the group velocity of light. There are many important applications, including slow light[4,5], optical delay lines[6], optical storage[7] and low-loss optical devices[8] so on. However, the experimental realization of EIT effect in atomic system usually requires some complicated conditions, such as refrigeration temperature, high intensity laser, etc. Therefore, the application of EIT effect in practice has been greatly limited.

In recent years, metamaterials have received widespread attention due to their significant applications in optical communications[9-11], bio-medicine sensing[12,13], and high-precision imaging[14-16]. Metamaterials are a class of artificial electromagnetic media, aiming to provide some controllable electromagnetic properties, such as room temperature conditions and large operating bandwidth in EIT-like effect[17-21]. The appearance of metamaterials provide a possibility to simulate EIT-like effect in classical optical systems[22].

In 2008, Zhang *et. al.* first proposed the concept of EIT-like metamaterials[23], they pointed out that the metamaterial can simulate the EIT-like effect by adding a special resonance structure (dark state resonance unit). Then, they used the near-field coupling principle between bright and dark resonant units to explain EIT-like effect. At the same time, Papasimakis *et. al.*[24] designed a multilayer fish-scale EIT-like metamaterial in order to broaden the working bandwidth in the microwave regime. Subsequently, researches successively proposed a series of improved metamaterial units, including square patch[25, 26], splitting resonant rings[27-29], cutting lines[30-32], U-shaped resonant structures[33-35] and other multilayer structures[36,37]. These works have proved theoretically and experimentally that the EIT effects can be realized through bright-dark mode coupling or bright-bright mode coupling in metamaterials[38,39]. However, most of the EIT-like metamaterials only have a single transparent window. Considering the multiple transparent windows have many important applications in multi-band filters and multi-band slow-light devices (such as delay lines[40,41]), in 2010, Kim employed bull's-eye structure to realize multiple transparent windows in terahertz band[42]. In 2012, Zhu *et. al.* proposed can realize the multiple transparent window EIT-like metamaterial and gave out slow light characteristics analysis by using different sizes cutting lines parallel to each other[30]. In 2018, Zhao *et. al.* proved that grating structure also can realize the EIT-like effect by controlling the grating period and the slot depth[43]. In this paper, we put forward a possiblity to realize multiple transparent windows in a grating structure metamaterial. The simulation results of electric field distributions show that the multi-band slow light effect can be obtained in the terahertz band. In addition, we need to consider the near-field bright-bright mode coupling mechanism between two U-shaped resonators with different resonant frequencies. Finally, we give out the three periods phase change of grating structure and the group delay. Thus, the multiple transparent windows have a potential application of the multi-band slow light.

## 2. The schematic diagram of grating structure metamaterial

The schematic diagram of grating structure metamaterial as shown in Figure 1. The three-period grating structure (3P-RL) unit etched on the top side of a $t1 = 20 \mu m$ thick polytetrafluoroethylene (PTFE) substrate. For the sake of simplicity, we will use 3P-RL to describe this structure in the following article. 3P-RL units structure size are chosen as $a = 40 \mu m, b = 30 \mu m, c = 4.5 \mu m, d = 4 \mu m$, respectively. In order to generate different resonance frequencies, we adopt three different slot depths, $e_1 = 35 \mu m, e2 = 30 \mu m, e3 = 25 \mu m$, respectively. The 3P-RL units is made of gold with a thickness of $t2 = 200\ nm$ and are periodically arranged with a lattice constant $P = 60 \mu m$ in both the $x$ and $y$ directions.

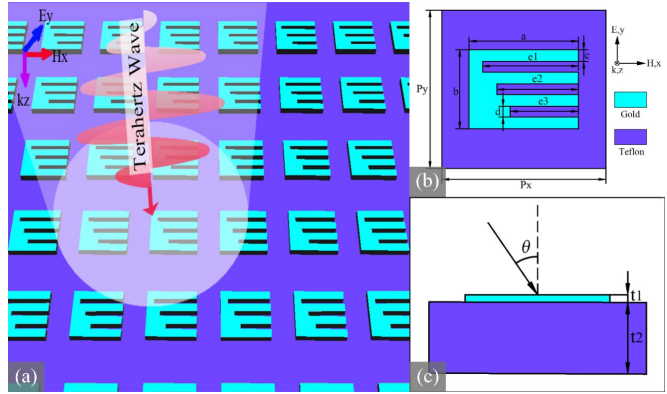

**Figure 1**. (a) Schematic diagram of the proposed 3P-RL metamaterial under the action of terahertz waves, the THz wave is TE polarization ($E \parallel y$ axis). (b) The top view of the gold-based unit cell. (c) The Side view of the gold-based unit cell. ($\theta$ is the angle between the incident THz wave and the normal of the material surface.)

In addition, Table 1 compares the frequency range and the simplicity of the structure to achieve multiple EIT effects. The table shows that the structure of our paper can be used with a relatively simple structure to achieve multiple EIT effects in the terahertz range.

**Table 1.** The comparison of the frequency range and the simplicity of the structure to achieve multiple EIT effects.

| Ref. | Frequency (THz) | Structure | Number of layer | Number of EIT-like windows ($n$) | Simplicity |
|---|---|---|---|---|---|
| [42] | 1~3.5 | multiple concentric microrings | Single | $n \geq 2$ | complicated |
| [30] | $6 \sim 13 \times 10^{-3}$ | Three different sizes cutting wires | Single | $n \geq 2$ | simple |
| [44] | $2 \sim 4 \times 10^{-3}$ | arc slot microring | Single | $n \geq 3$ | complicated |
| [36] | 3~5 | Graphene/dielectric multilayer periodic array | multiple | $n \geq 3$ | complicated |
| This work | 0.9~2.2 | grating-like | Single | $n \geq 2$ | simple |

The Frequency Domain Finite Integration Method (CST Microwave Studio) is used for the normal incident angle ($\theta = 0°$) of the electromagnetic wave with periodic boundary conditions of the X and Y directions and open space for the Z direction. The relationship between this angle and the transmission spectrum is described in detail in Figure.2

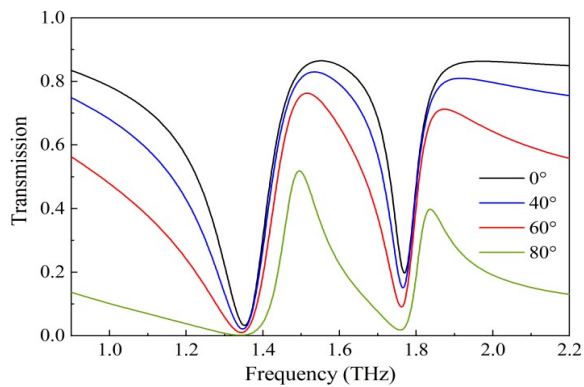

**Figure 2**. The relationship between oblique incidence and frequency of 2P-RL units metamaterials. $\theta$ is the angle between $k$ and zaxis: $\theta = 0°$(black line), 20°(red line), 40°(blue line), 60°(green line), 80°(purple line). The parameters used in the simulated transmission spectrum in the figure are the optimized 2P-RL parameters.( $a = 40\mu m, b = 30\mu m, c = 4.5\mu m, d = 4\mu m, e1 = 35\mu m, e2 = 27.1\mu m, t1 = 20\mu m, t2 = 200nm$)

The transmissive spectrum of the metamaterial composed of two cycles of each incident angle as shown in Figure 2. With the increasing of the incident angle, the transmission peak gradually decreases. Especially at $\theta = 80°$, the transmission peak attenuate to 0.7. In summary, the EIT-like effect is the best at normal incidence ($\theta = 0°$).

A medium grid size is used to ensure simulation accuracy. In the numerical simulation, the dielectric constant of the PTFE substrate is $\varepsilon = 2.2$ [30]. Gold has the following optical constants described by the Drude model at terahertz frequencies[45]:

$$\varepsilon_{Gold} = \varepsilon_\infty - \frac{\omega_p^2}{\omega^2 + i\omega\gamma} \quad (1)$$

where the plasmon frequency $\omega_p = 1.37225 \times 10^{16} rad/s$ and the damping constant $\gamma = 4.0527 \times 2\pi \times 10^{13} rad/s$.

## 3. Results and discussions

In order to clarify the physical mechanism of EIT-like resonance, we first study the metamaterial composed of two-period grating structure (2P-RL) units. For 3P-RL units, we adopt the similar method.

As shown in Figure 3, the slot depth difference $\Delta e = e1 - e2$ will affect the bandwidth of EIT-like transmission spectrum and the resonance window transmittance difference.

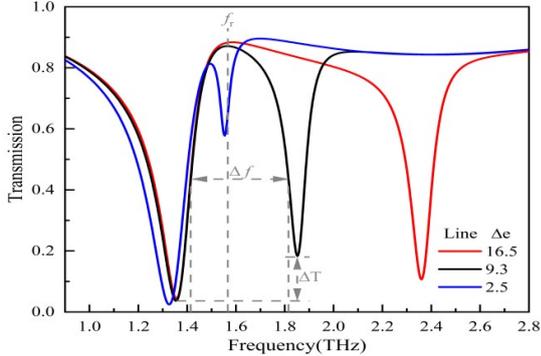

**Figure 3.** EIT-like transmission spectra of metamaterials composed of 2P-RL units when the slot depth difference $\Delta e$ is equal to $2.5\mu m$(blue line), $9.3\mu m$(black line) and $16.5\mu m$(red line), respectively. $\Delta f$ is the bandwidth, $\Delta T$ is the transmittance difference of the resonance window, and $f_r$ is the resonance frequency. Here, $e1 = 35\mu m$, $e2 = 18.5 \sim 35\mu m$.( In order to make the picture look concise, only three $e2$ values are shown here.)

Next, we use the $Q$ factor to get the optimized $\Delta e$:

$$Q = \frac{f_r}{\Delta f} \quad (2)$$

where $f_r$ is the resonance frequency and $\Delta f$ is the full width at half maximum (FWHM) of the resonance window[31].

According to Eq.(2), we can give out the relationship between the slot depth difference $\Delta e$, the FWHM $\Delta f$, the resonance window transmittance difference $\Delta T$ and the quality factor $Q$. Figure 4 clearly shows that $\Delta T$ and $Q$ are negatively correlated with $\Delta e$, while $\Delta f$ is positively correlated with $\Delta e$.

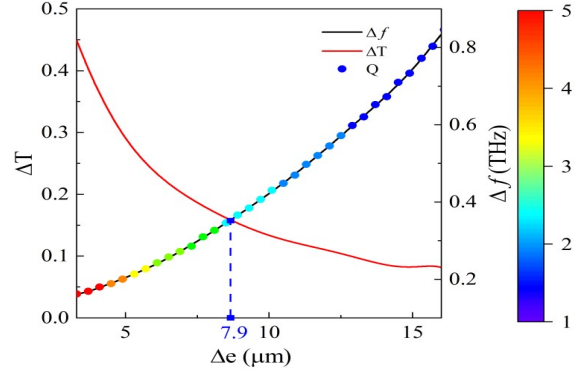

**Figure 4.** As the slit depth difference $\Delta e$ increases, the changes in $\Delta f$ (the bandwidth, black line), $\Delta T$ (the difference in transmittance of the resonance window, red line), and $Q$ factor(point).

We hope that the smaller the resonance window transmittance difference $\Delta T$, the larger the quality factor $Q$, so we can get an optimal value $\Delta e = 7.9\mu m$ in the Figure 4.

Then, by optimizing the value of $\Delta e$, we can obtain the EIT-like transmission spectrum of the metamaterial composed of 2P-RL units, as shown in Figure 5. In terms of Figure 5(a), we can observe that 2P-RL units have two resonance peaks at 1.35 THz(I) and 1.77 THz(III), and a transparent window at 1.55 THz(II). The theoretical predictions results are shown to agree well with the simulated transmission spectra. Here, we regard 2P-RL units as two U-shaped resonators. The upper part of the 2P-RL units resonate at 1.35THz, while the lower part of the 2P-RL units resonate at 1.77THz. The two resonators mutual coupling results in a transparent window at 1.55 THz. It can be seen from Figure 5(b) that the electric field distribution diagram at 1.35THz, 1.55THz and 1.77THz, respectively: At 1.35THz and 1.77THz, due to the resonator generated by 2P-RL units, the energy of the incident wave is concentrated on the top of the two U-shaped resonators.

At 1.55THz, due to the two U-shaped resonators coupling, the transparent window allows most of the energy of the incident wave to move through.

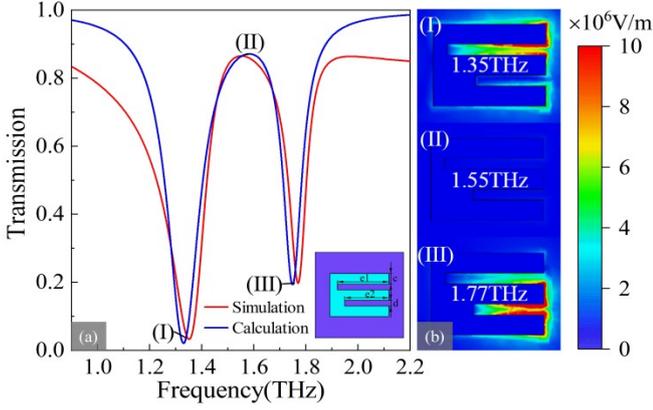

**Figure 5.** (a) EIT simulation(red line) and calculated(blue line) transmission spectrum of two-period raster-like (2P-RL). Here, $e1 = 35\mu m, e2 = 27.1\mu m$; (b) The spatial distribution of the resonant electric field $E_z$ in a single unit cell composed of 2P-RL, calculated at 1.35 THz, 1.55 THz, and 1.77 THz, respectively.

We clarify the EIT-like effect theoretically are based on the classic two-particle model. In the 2P-RL units, we regard the two U-shaped resonators as two particles, namely particle a and particle b. The near-field coupling between two particles under incident linearly polarized terahertz waves can be analytically described as[46]:

$$\ddot{x}_a(t) + \gamma_a \dot{x}_a(t) + \omega_a^2 x_a(t) + \kappa^2 x_b(t) = \frac{q_a}{m_a} E,$$
$$\ddot{x}_b(t) + \gamma_b \dot{x}_b(t) + \omega_b^2 x_b(t) + \kappa^2 x_a(t) = \frac{q_b}{m_b} E. \quad (3)$$

Here, $q_a$ and $q_b$, $m_a$ and $m_b$, $\omega_a$ and $\omega_b$, $\gamma_a$ and $\gamma_b$ are denoted as the effective charge, effective mass, resonance angular frequency, and loss factor of particle a and particle b, respectively. $\kappa$ is defined as the coupling coefficient between two particles. Here, we consider the two particles interaction with the incident terahertz electric field $E = E_0 e^{i\omega t}$, giving rise to the EIT-like destructive interference between bright mode and dark mode. Now, by expressing the displacement vector of particle a and b as $x_a = ae^{-i\omega t}$ and $x_b = be^{-i\omega t}$, respectively, we can solve the Eqs. (1) and (2) for $x_a$ and $x_b$:

$$x_a = \frac{m_a q_b \kappa^2 + m_b q_a B}{m_a m_b (\kappa^4 - AB)} E,$$
$$x_b = \frac{m_b q_a \kappa^2 + m_a q_b A}{m_a m_b (\kappa^4 - AB)} E. \quad (4)$$

Where $A = (\omega^2 - \omega_a^2 + i\omega\gamma_a); B = (\omega^2 - \omega_b^2 + i\omega\gamma_b)$. Next, we calculate the polarization rate $\chi$ by the relationship between the polarization rate $\chi$, the polarization intensity $P$, and the incident field $E$:

$$\chi = \frac{P}{\varepsilon_0 E} = \frac{q_a x_a + q_b x_b}{\varepsilon_0 E}$$
$$= \frac{Am_1 q_2^2 + Bm_2 q_1^2 + q_1 q_2 \kappa^2 (m_1 + m_2)}{m_1 m_2 \varepsilon_0 (AB - \kappa^4)} \quad (5)$$

The real and imagery part of $\chi$ indicates the dispersion and absorption in the structure, respectively. Based on the conservation of energy relation $T + A = 1$ (normalized to unity), where $A = Im(\chi)$ is the absorption(loss). The transmission spectra $T = 1 - Im(\chi)$ (given by the Kramers-Kronig relations) can be used to fit data in the simulated transmission spectra (see Figure. 5(a)).

We plot the transmission spectrum of the two-particle model in Figure 5(a), amplitude and bandwidth has a slight difference from the numerical simulation curves. The slight difference may be results from the dispersion.

Next, we generalize the 2P-RL units situation to 3P-RL units situation, and obtain the EIT-like transmission spectrum and electric field distribution through numerical simulation. We can generate multiple transparent windows by increasing the number of U-shaped resonators, that is, the period of the unit structure.

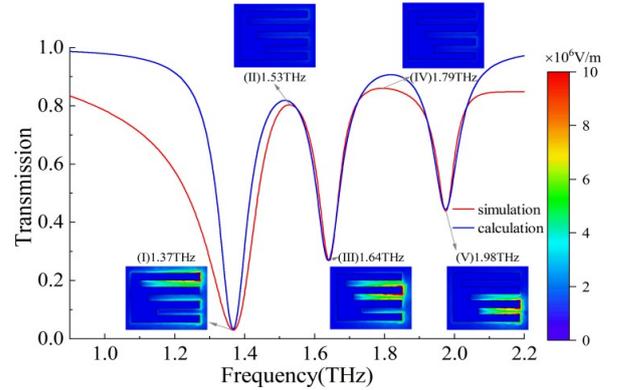

**Figure 6.** The multiple EIT-like transmission spectrum of 3P-RL units and the electric field spatial distribution by numerical simulation (the red line) and theoretical calculation (the blue line) at 1.37 THz, 1.53 THz, 1.64 THz, 1.79 THz and 1.98 THz, respectively.

In terms of Figure 6, we can see that 3P-RL units can produce resonance peaks at 1.37 THz (I), 1.64 THz (III) and 1.98 THz (V), and create transparency window at 1.53 THz (II) and 1.79 THz (IV). Here, we regard 3P-RL units as three U-shaped resonators. The three U-shaped resonators corresponds to the three resonance peaks, respectively. The two transparent windows are formed based on the near-field coupling mechanism. This can be seen from the electric field distribution diagram, in that almost all the resonance peak energy is concentrated on the top of the U-shaped resonators when the resonance occurs.

We consider theoretically the three U-shaped resonators as three particles coupling model. Under the incident linearly polarized terahertz wave, the near-field coupling effect between the three particles can be analytically described as:

$$\ddot{x}_a(t) + \gamma_a \dot{x}_a(t) + \omega_a^2 x_a(t) + \kappa_{ab}^2 x_b(t) = \frac{q_a}{m_a} E,$$
$$\ddot{x}_b(t) + \gamma_b \dot{x}_b(t) + \omega_b^2 x_b(t) + \kappa_{ab}^2 x_a(t) + \kappa_{bc}^2 x_c(t) = \frac{q_b}{m_b} E, \quad (6)$$
$$\ddot{x}_c(t) + \gamma_c \dot{x}_c(t) + \omega_c^2 x_c(t) + \kappa_{bc}^2 x_b(t) = \frac{q_c}{m_c} E.$$

Here, $q_a$, $q_b$, $q_c$; $m_a$, $m_b$, $m_c$; $\omega_a$, $\omega_b$, $\omega_c$; $\gamma_a$, $\gamma_b$, $\gamma_c$ are denoted as the effective charge, the effective mass, the resonance angular frequency, and the loss factor of particle a, b and c, respectively. $\kappa_{ab}$ and $\kappa_{bc}$ are the coupling coefficients of particle a, particle b and particle c, respectively, giving rise to the multiple EIT-like destructive interference between bright mode and dark mode. The same can be obtained by using $\chi = P/\varepsilon_0 E$.

$$\chi = \frac{1}{N}\{ACm_am_cq_b^2(A(C - q_c\kappa_{bc}^2)) - Cq_a\kappa_{ab}^2 + m_b[m_aq_c(Cq_b\kappa_{bc}^2(q_a\kappa_{ab}^2 - A) + q_c(C(AB - \kappa_{ab}^4) + A\kappa_{bc}^4(q_b - 1))) + Cm_cq_a(Aq_b\kappa_{ab}^2(q_c\kappa_{bc}^2 - C) + q_a(A(BC - \kappa_{bc}^4) + C\kappa_{ab}^4(q_b - 1)))]\}$$
(7)

Here $N = ACm_am_bm_c\varepsilon_0(ABC - C\kappa_{ab}^4 - A\kappa_{bc}^4)$; and $A = (\omega^2 - \omega_a^2 + i\gamma_a\omega)$; $B = (\omega^2 - \omega_b^2 + i\gamma_b\omega)$; $C = (\omega^2 - \omega_c^2 + i\gamma_c\omega)$; we use $T = 1 - Im(\chi)$ to obtain the theoretical EIT-like transmission spectrum under the three-particle model (3P-RL units).

A notable application of EIT-like is the ability to obtain slow light through optical communication. Group delay represents the time delay of narrow-band optical pulses in optical devices. The group delay $\tau_g$ is defined in optical elements as the differential of negative spectral phase versus frequency $\tau_g = -d\varphi(f)/df$ [31], where $f$ is the frequency. The phase data in the figure is obtained by simulation. We can observe in Figure 7 that the strong phase dispersion around the transparent window causes a large $\tau_g$. In the vicinity of the transparency peak, large positive group delays are obtained, indicating potential use in slow light applications.

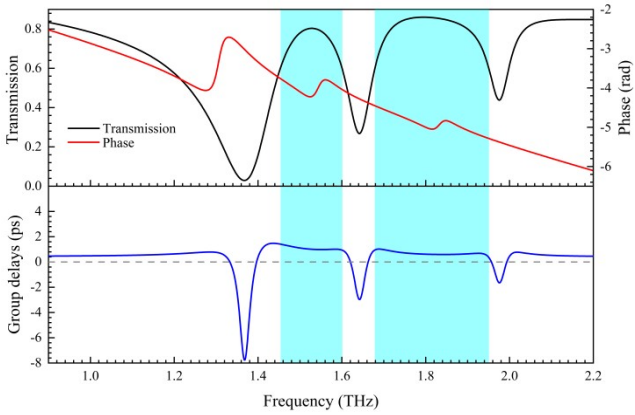

**Figure 7.** Phase change of 3P-RL units and the group delay at 0.9THz to 2.2THz.

## 4. Conclusion

In conclusion, we have demonstrated the transmission spectra of two and three-period grating-like unit structures through numerical simulations and theoretical calculations and analysed the causes of resonance peaks and transparent windows by studying its electric field distribution-the unit structure can be seen as multiple U-shaped resonators, and the near-field coupling effect between them leads to the EIT-like effects in the terahertz band. The larger group delay $\tau_g$ caused by the EIT-like effects can be applied in slow light. The influence of the change of incident angle $\theta$ on the transmission spectrum is shown in Figure 7. We choose vertical incidence (($\theta = 0°$) to obtain the most optimized transmission spectrum. As we can see in this paper, we can generate multiple transparent Windows by increasing the number of U-shaped resonators. Therefore, it can be applied in multi-band filters and multi-band slow light devices (such as delay lines).

## Acknowledgments

This work was supported by the National Natural Science Foundation of China (11504074) and the State Key Laboratory of Quantum Optics and Quantum Optics Devices, Shanxi University, Shanxi, China (KF202004).